\documentclass[twocolumn,aps]{revtex4}
\usepackage{amsmath}
\usepackage{amssymb}
\usepackage{graphicx}
\usepackage{dcolumn}
\usepackage{bm}
\usepackage{xcolor}
\usepackage{epstopdf}

\begin{document}

\title{Multiple crossing of Landau levels of two-dimensional fermions in
double HgTe quantum wells}

\author{G. M. Gusev,$^1$ E. B. Olshanetsky,$^{2}$ F. G. G. Hernandez,$^1$
O. E. Raichev,$^3$ N. N. Mikhailov,$^2$ and S. A. Dvoretsky$^2$}

\affiliation{$^1$Instituto de F\'{\i}sica da Universidade de S\~ao
Paulo, 135960-170, S\~ao Paulo, SP, Brazil}
\affiliation{$^2$Institute of Semiconductor Physics, Novosibirsk
630090, Russia}
\affiliation{$^3$Institute of Semiconductor Physics, NAS of
Ukraine, Prospekt Nauki 41, 03028 Kyiv, Ukraine}

\begin{abstract}

The double quantum well systems consisting of two HgTe layers separated by a tunnel-transparent barrier
are expected to manifest a variety of phase states including two-dimensional gapless semimetal and
two-dimensional topological insulator. The presence of several subbands in such systems leads to a rich
filling factor diagram in the quantum Hall regime. We have performed magnetotransport measurements of
the HgTe-based double quantum wells in both gapless and gapped state and observed numerous crossings
between the Landau levels belonging to different subbands. We analyze the Landau level crossing
patterns and compare them to the results of theoretical calculations.

\end{abstract}

\maketitle
\section{Introduction}

The energy band structure of two-dimensional (2D) carriers in HgTe quantum wells (QWs) \cite{konig} is described
by the Bernevig-Hughes-Zhang (BHZ) effective Hamiltonian \cite{bernevig}, which is similar to the Dirac Hamiltonian
for 2D fermions. This description accounts for intriguing properties observed experimentally, such as the
existence of two-dimensional topological insulators (TI) with edge states \cite{konig, hasan, qi, kane,
moore, moore2, zhou, roth, gusev}, massless Dirac fermions with band dispersion similar to that of graphene
\cite{buttner, kozlov, gusev2}, and phase transitions between the ordinary insulator and the 2D TI. The double
quantum wells (DQWs) consisting of two tunnel-coupled HgTe QWs, Fig. 1 (a), represent the simplest example of
multilayer Dirac fermion system with variable properties. Such DQWs have been recently recognized as promising systems
to explore the phase transition between topological and ordinary insulator states \cite{michetti, kristopenko}.
The additional degrees of freedom associated with tunnel-induced hybridization of different 2D subbands lead
to a variety of possible phase states. The conditions for subband inversion corresponding to transitions
between these states depend on the well width $d$ and on the coupling parameters, such as the thickness
$t$ and height of the potential barrier separating the QWs \cite{kristopenko}. The studies of HgTe-based
DQWs are of particular interest in connection with possible realization of TIs with tunable gaps.

\begin{figure}[ht!]
\includegraphics[width=8.5cm,clip=]{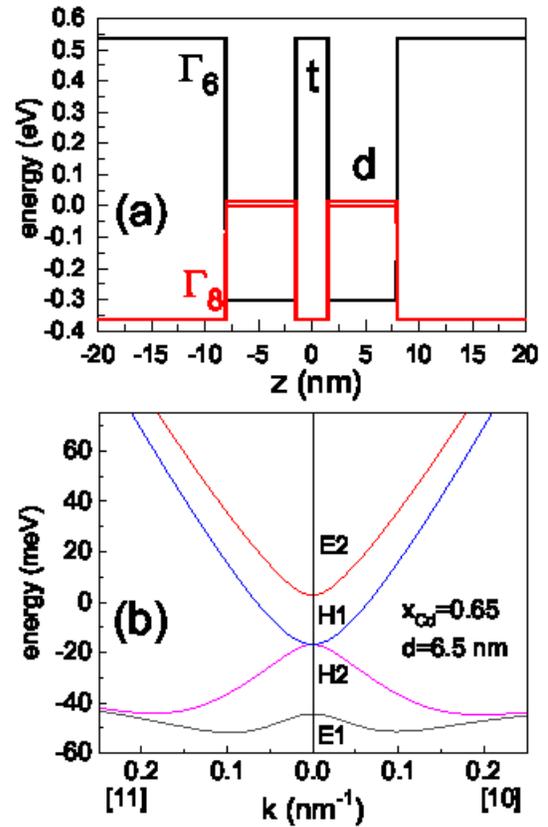}
\caption{\label{fig.1}(Color online) (a) Band profile of the symmetric [013]-grown HgTe DQW with
$d=6.5$ nm, $t=3$ nm, and Cd content $x_{Cd}=0.65$. (b) Energy dispersion of 2D subbands along two
directions of the wave vector ${\bf k}$ in the growth plane $xy$, $k_y=0$ ([10]) and $k_y=k_x$ ([11]).}
\end{figure}

The diversity of phase states in HgTe DQWs is related to a negligible tunnel hybridization of the heavy hole subbands
H1 and H2 at zero wave vector ${\bf k}$, so these subbands remain degenerate, while the electronlike subbands E1
and E2 experience a strong hybridization leading to a large energy separation between them. As a result, different
kinds of subband ordering can occur. The fully inverted subband ordering E1-E2-H2-H1 naturally corresponds to the TI
phase. On the other hand, the "mixed" subband ordering E1-H2-H1-E2, which can be realized in a wide region of parameters,
leads to the gapless spectrum in symmetric DQWs, as shown in Fig. 1 (b). This gapless spectrum resembles that of
bilayer graphene \cite{castro}, but with a crucial difference, the absence of the valley degeneracy. However, if
the symmetry is broken by a transverse bias or electric field perpendicular to the 2D plane, the heavy hole subbands
H1 and H2 become split apart. Thus, similar to the case of bilayer graphene \cite{castro,mocann}, a transverse
electric field transforms the gapless phase into a gapped (insulating) phase \cite{kristopenko}. In contrast to
bilayer graphene, where the gapped state is an ordinary insulator, the gapped state in HgTe DQWs is expected to
be a 2D TI. Indeed, the inverted subband ordering, when E1 falls below H1 and H2, leads to coexistence of the bulk
and edge states, and when the gap opens in the bulk, the edge states remain gapless in view of the time reversal
symmetry. Recent studies \cite{gusevnew} confirm the existence of the gap and edge state transport both in symmetric
and asymmetric DQWs, as well as the existence of the gapless phase, in accordance with theoretical predictions.

Application of a magnetic field perpendicular to the layers leads to Landau quantization and allows one to get
additional information about electronic properties of the systems with Dirac-like band spectrum. In single layer systems,
it makes possible to distinguish between the gapless phase with Dirac cone dispersion (graphene and HgTe QWs) and
gapped (TI and ordinary insulator) phases in HgTe QWs. In this case, there exists a single crossing of special Landau
levels (LLs) with zero indices ($n=0$) belonging to conduction and valence subbands, which takes place in the TI phase
\cite{konig, castro, peres, sarma, buttner}. In contrast to single layer systems, the LL spectrum both in bilayer
graphene and in HgTe-based DQWs is expected to show numerous LL crossings \cite{falko, nilsson, kristopenko, koshino, fogler}.
It is found that multiple LL crossing provides a useful tool for precise determination of the band structure parameters in
trilayer graphene \cite{herrero, stepanov, datta}. In bilayer graphene, observation of such crossing is unlikely because of
large separation of subbands in energy space \cite{chung}. On the other hand, in HgTe-based DQWs the subband
energy separation is small (Fig. 1), which makes them unique, among the other systems with Dirac-like energy
spectrum, for experimental studies of multiple LL crossing. Unlike the case of graphite multilayers, the
absence of valley degeneracy in the HgTe structures allows for a completely reliable comparison between the
theoretical LL spectrum and the experimentally observed LL crossing and splitting. Studies of the properties
of energy spectrum of multilayer Dirac systems in magnetic field are important also in connection with a growing
interest to the quantum Hall ferromagnetism \cite{datta} and multicomponent fractional quantum Hall effect (QHE)
\cite{papic}.

Though experimental studies of HgTe DQWs in magnetic field are limited \cite{yakunin,bovkun}, they already
demonstrate some features related to complex LL spectrum in these structures. In particular, studies of the QHE in
the samples with hole type of conductivity reveal a reentrant quantization of Hall resistance \cite{yakunin}.
Magnetooptical spectra show shifts and doubling of the main magnetoabsorption lines \cite{bovkun}. These features
are in qualitative agreement with LL spectrum calculation based on the Kane Hamiltonian. However, the magnetotransport
measurements \cite{yakunin} have been carried out in a narrow interval of carrier density, changed by illumination.
While showing important aspects of the Landau quantization in the valence band, the results of Ref. \cite{yakunin}
do not give a comprehensive picture of LL spectrum. In particular, the behavior of LLs in the conduction band, where
a regular pattern of multiple LL crossing is expected, and in the important region near the charge neutrality remains
unexplored.

In this paper, we use DQW samples with a top gate which allows us to control the density of carriers in a wide interval
and to determine the LL picture from magnetotransport measurements. We study two kinds of HgTe-based double quantum wells
with different values of $x_{Cd}$ and $d$. The DQW with $d=6.5$ nm, whose band diagram and energy spectrum are shown in Fig. 1,
demonstrates the properties of gapless semimetal, while the DQW with $d=6$ nm shows an activation temperature dependence of
conductivity corresponding to a small energy gap $\Delta=11$ meV \cite{gusevnew}. As the symmetric DQW with $d=6$ nm
should also be in a gapless phase according to our calculations, the presence of the gap can be explained by an asymmetry,
for example, the appearance of an internal electric field due to uncontrollable doping during the growth. For both DQWs,
we have investigated the LL diagram and revealed multiple successive crossings between LLs in magnetic fields up to 8 T,
resembling those observed in trilayer graphene \cite{herrero, stepanov, datta}. For DQW with $d=6.5$ nm, we have studied
behavior of the resistance near the charge neutrality point, which identifies the important crossing point for the LLs
with $n=0$ and suggests that the degeneracy of H1 and H2 subbands is lifted as a result of application of the gate voltage.
The observed properties of the energy spectrum are reproduced in our numerical calculations.

The paper is organized as follows. In Sec. II we provide the details of the sample fabrication and measurements, and
describe the details of the calculations. In Sec. III we present the results of measurements of both the longitudinal
and the Hall resistances in magnetic fields, compare them with similar measurements in single quantum wells, and discuss
these results in correlation with the calculated LL spectra. The last section contains a brief summary and conclusions.

\section{Methods}

The structures containing HgTe quantum well layers of equal widths separated by Cd$_{x}$Hg$_{1-x}$Te barriers,
with surface orientation [013], were grown by molecular beam epitaxy (MBE). To prepare the gate, a dielectric
layer (100 nm of SiO$_{2}$ and 100 nm of Si$_{3}$ Ni$_{4}$) was first grown on the structure using the
plasmochemical method, and then the TiAu gate was deposited. By studying two groups of samples with different
parameters $x_{Cd}$ and $d$, we have observed both the gapped TI and the gapless semimetal phases, distinguished
by the temperature dependence of the local resistance. Table I contains the list of experimental samples and
indicates the typical parameters, such as the well width $d$, barrier thickness $t$, barrier composition
$x_{Cd}$ and gate voltage corresponding to the charge neutrality point (CNP) $V_{CNP}$. By measuring nonlocal
resistance in zero magnetic field in the gapped DQW samples with $d=6$ nm, we confirm the presence of the edge
state transport in the absence of magnetic field \cite{gusev4}, thereby providing the experimental evidence
of the 2D TI phase in HgTe DQWs \cite{gusevnew}. Neither the activation behavior of the conductance nor the
edge state transport are observed in the DQWs with $d=6.5$ nm, expected to be in the gapless phase \cite{gusevnew}.

\begin{table}[ht]
\caption{\label{tab1} Parameters of the HgTe/Cd$_{x}$Hg$_{1-x}$Te DQWs at $T=4.2$ K.}
\begin{ruledtabular}
\begin{tabular}{lccccc}
$d$ (nm) & $t$ (nm) & Cd content $x$ & $V_{CNP}$ (V)& properties\\
\hline
6.5 & 3 & 0.65 & -4.2 & gapless \\
\hline
6.0 & 3& 0.37 & -0.7& gapped TI \\
\hline
\end{tabular}
\end{ruledtabular}
\end{table}

The quantum well width and the thickness of the barrier $t=3 \pm 0.3$ nm have been determined by using
ellipsometry during the MBE growth. The samples are Hall bar devices with 2 current probes and 7 voltage
probes. The bar has a width $w=4$ $\mu$m and three consecutive segments of different lengths $L=2$, 8,
and 32 $\mu$m (Fig. 2). Fabrication of ohmic contacts to HgTe quantum well is similar to that for other 2D systems,
such as GaAs quantum wells: the contacts were formed by the burning of indium directly to the surface of large contact
pads. The modulation-doped HgTe/CdHgTe quantum wells are typically grown at 180$^{\rm o}$ C. Therefore, in contrast
to III-V compounds, the temperatures during the contact fabrication process are relatively low. On each
contact pad, the indium diffuses vertically down, providing an ohmic contact to both quantum wells, with the
contact resistance in the range of 10-50 kOhm. During the AC measurements we always check that the Y-component
of the impedance never exceed 5\% of the total impedance, which demonstrates good ohmicity of the contacts.
Three devices for each set of parameters were studied. The density variation with the gate voltage was
$0.9 \times 10^{11}$ cm$^{-2}$V$^{-1}$. The magnetotransport measurements were performed in the temperature range
$1.4-70$ K by using a standard four-point circuit with a $1-27$ Hz ac current of $1-10$ nA through the sample,
which is sufficiently low to avoid overheating effects. It is worth noting that in the presence of strong magnetic
fields there may occur a resistance build-up in the mesa across the gate electrode edge due to carrier
reflection. This is why we report the results only up to 4 T or 8 T, depending on the device.
The current flows through the contacts 1 and 5, the longitudinal voltage is measured between the closest pair
of voltage probes (9 and 8), and the Hall voltage is measured between the probes on the opposite sides of
the Hall bar.

The numerical calculations of electron energy spectrum were based on the $6 \times 6$ Kane
Hamiltonian whose description with applications for different orientations of the heterointerface
is given in Ref. \cite{raichev}. The energy is counted from the $\Gamma_8$ band edge in HgTe, the
relative band edge positions of the $\Gamma_6$ in HgTe, $\Gamma_6$ in CdTe, and $\Gamma_8$ in CdTe are
$-0.3$ eV, $0.99$ eV, and $-0.56$ eV, respectively. The Kane matrix element $P$ and the Luttinger
parameters $\gamma_{1}$, $\gamma_{2}$, and $\gamma_{3}$ for HgTe and CdTe are taken from
Ref. \cite{novik}. Also, a strain-induced shift $\varepsilon_s = 22$ meV of light hole band with
respect to heavy hole band in HgTe layer on CdTe substrate \cite{brune} has been taken into
account. Applying these parameters for the structures with Cd$_x$Hg$_{1-x}$Te barriers, we use
a linear interpolation with respect to Cd content in order to find the band edge energies,
Luttinger parameters in the barriers, and $\varepsilon_s$ in HgTe layers. The energy spectra at
zero $B$, shown in Fig. 1, were calculated for [013] orientation of the interface. The LL spectra
shown in Fig. 5 (a) and Fig. 6 (a) were calculated in the isotropic approximation, when $\gamma_{3}$
was set equal to $\gamma_2$ both in the wells and in the barrier layers (in this case the spectra
do not depend on the orientation). The isotropic approximation strongly simplifies the numerical
solution of eigenstate problem \cite{raichev1} and is expected to give reliable results for the
conduction band spectrum, which is in the focus of attention in this paper.

\begin{figure}
\includegraphics[width=8cm,clip=]{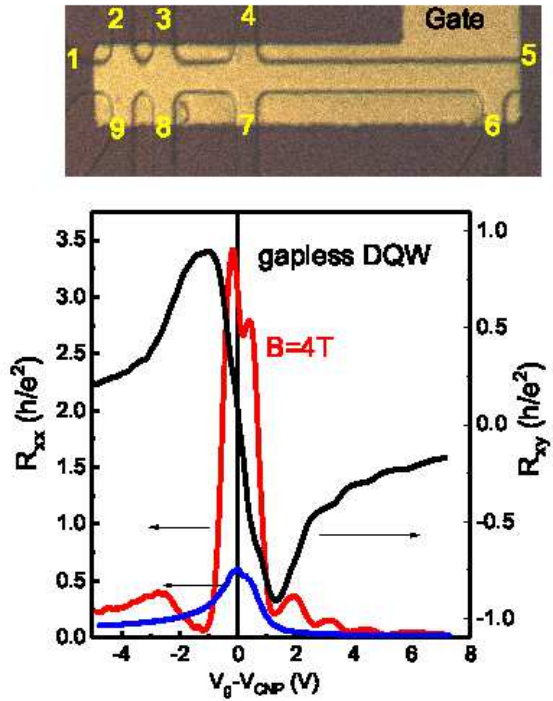}
\centering
\caption{\label{fig.2}(Color online) Zero-field resistance (blue), longitudinal (red) and Hall (black)
resistances at $B=4$ T and $T=4.2$ K as functions of the gate bias for the gapless ($d=6.5$ nm,
$x_{Cd}=0.65$) double quantum well. The top panel shows the device layout and numbering of the probes.}
\end{figure}

\section{Results}

In both kinds of DQW samples we study, the dependence of the resistance on the gate voltage in zero magnetic field
(Fig. 2) shows a maximum near the CNP, resembling the one observed in the HgTe-based single quantum wells,
including TIs \cite{konig, roth, gusev2} and 6.3 nm-wide QWs with massless Dirac fermions \cite{buttner, kozlov, gusev}.
In order to probe the nature of the transport in the DQWs, we have measured the temperature dependence of
the conductance near the CNP. The results are briefly described below. A detailed description can be found
in Ref. \cite{gusevnew}. While in narrow-gap samples one would expect a thermally activated conductance
(see, for example, a review paper \cite{gusev4}), the DQW with $d=6.5$ nm shows a quasi-metallic behavior,
the resistance is saturated for $T < 10$ K and increases with $T$ at higher temperatures. A similar
temperature dependence has been observed for the gapless Dirac fermion system in a single 6.3 nm HgTe
well \cite{buttner,kozlov,gusev}. The other DQW with lower Cd content, $x=0.37$, $d=6$ nm, and $t=3$ nm, shows
an insulating temperature dependence described by the activation law $R \propto \exp (\Delta/2k_B T)$ at $T > 15$ K.
The thermally activated behavior corresponds to a gap of energy $\Delta=11$ meV between the conduction and
the valence bands. This finding can be explained by the presence of an asymmetry, presumably related to the
specifics of the MBE growth for this particular structure, such as an uncontrollable chemical doping. Our
calculations demonstrate that the observed gap can be created by a transverse electric field $E=13$ kV/cm.
The results of earlier experiments on HgTe DQWs \cite{yakunin,bovkun} also suggest the presence of transverse
electric field. While in bilayer graphene the transverse electric field can be controlled by double gates
\cite{maher}, the MBE-grown HgTe DQWs with controllable gap are still not available technologically.

The measurements of longitudinal $R_{xx}$ and the Hall $R_{xy}$ resistances in a perpendicular magnetic field
demonstrates the QHE behavior: the resistance minima are accompanied with the Hall resistance plateaux (Fig. 2).
We first compare the magnetotransport in the gapless DQW with that in the gapless single QW. Figures 3 and 4 shows
the longitudinal $\sigma_{xx}$ and Hall $\sigma_{xy}$ conductivities as functions of both the gate voltage
and the magnetic field for these two structures. At the CNP, the longitudinal conductivity drops to zero, while the Hall
one changes its sign in both kinds of structures. The quantum Hall state with zero filling factor $\nu=0$ differs from
the states with $\nu \neq 0$. Indeed, for the states with $\nu \neq 0$ both the longitudinal conductivity
$\sigma_{xx}$ and the longitudinal resistivity $\rho_{xx}$ turn to zero, while for the QHE state $\nu = 0$
the longitudinal resistance shows a maximum rather than a minimum, $\rho_{xx} \gg \rho_{xy}$, and we obtain
$\sigma_{xx} \sim \sigma_{xy}$ \cite{abanin}. Early experiments on the quantum Hall effect in bilayer graphene
have confirmed an important theoretical prediction \cite{falko} that the nearest to the CNP Hall conductivity
plateau has the value $4e^{2}/h$, which is two times larger than that in a single layer graphene due to the
additional degeneracy of the zeroth Landau level. However, in high-mobility samples in a strong magnetic
field a complete symmetry breaking takes place and the QHE states appear at all integer filling factors
\cite{velasco,maher,weitz,lee}. In our HgTe DQW samples, the Hall conductivity plateaux are also observed
at all integer filling factors, because the valley degeneracy is absent and the Zeeman splitting is large
enough. While $\sigma_{xx}$ in a single QW shows regular peaks and minima, $\sigma_{xx}$ in DQWs is less
regular and shows modulations of Shubnikov-de Haas oscillations because of superposition of the contributions
from LLs belonging to different 2D subbands.

\begin{figure}[ht!]
\includegraphics[width=8.cm,clip=]{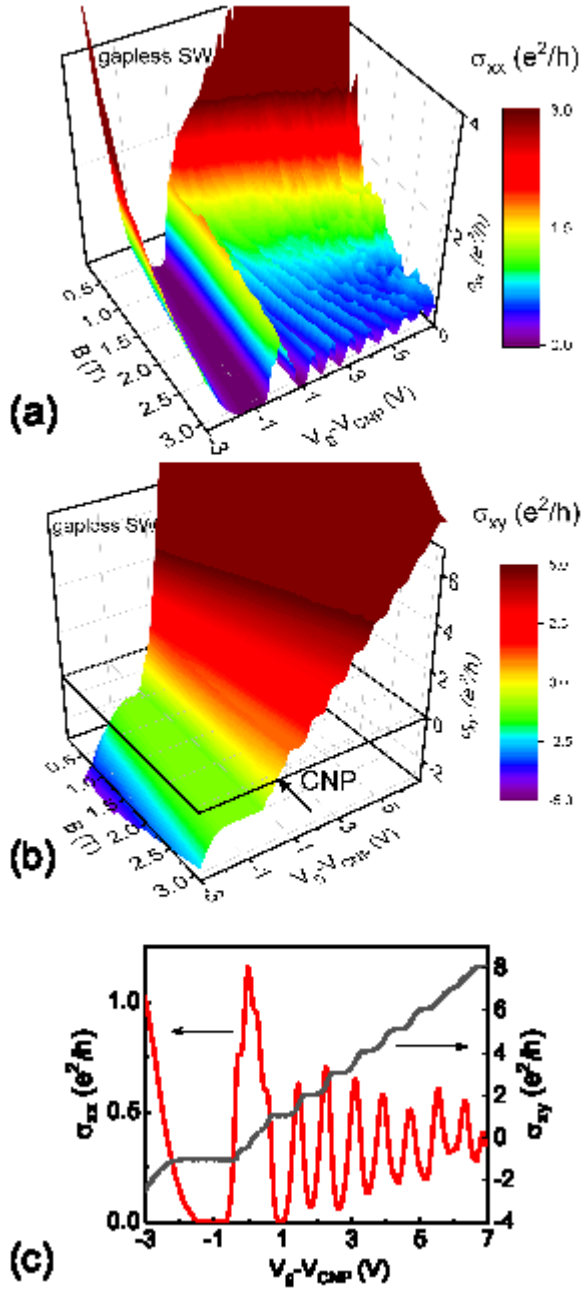}
\caption{\label{fig.3}(Color online)
The conductivities $\sigma_{xx}$ (a) and $\sigma_{xy}$ (b) as functions of the gate voltage and magnetic
field for a single quantum well with Dirac cone energy spectrum, and traces of $\sigma_{xx}$ and
$\sigma_{xy}$ at $B=3$ T (c).}
\end{figure}

\begin{figure}[ht!]
\includegraphics[width=8.cm,clip=]{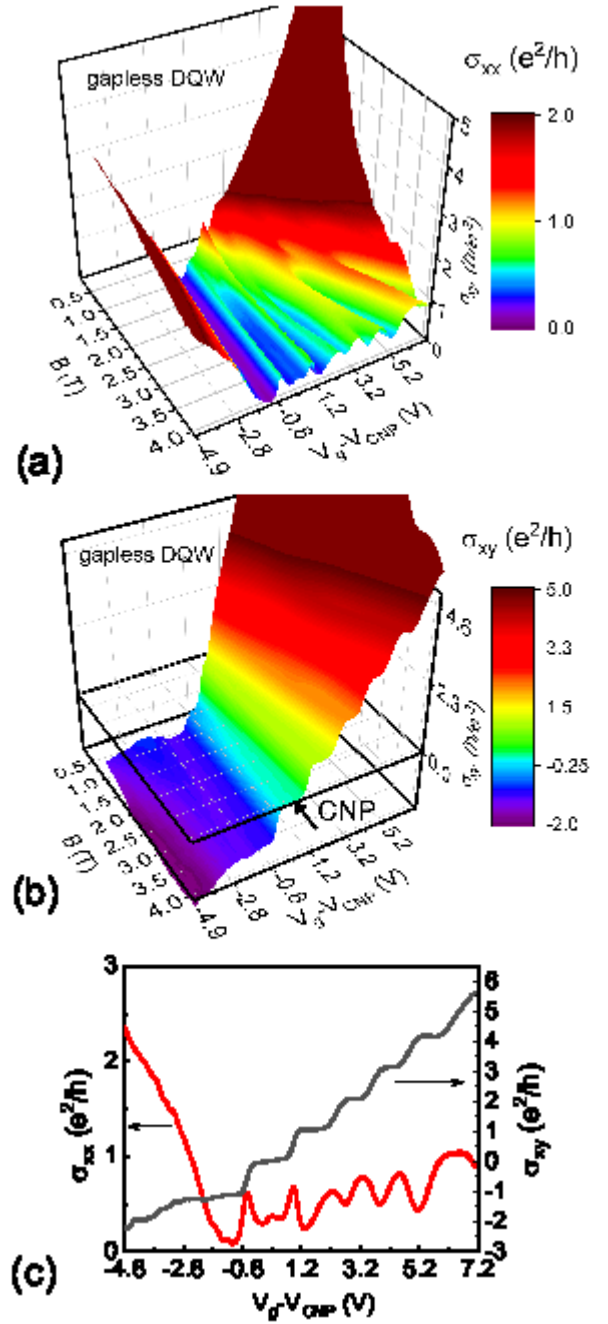}
\caption{\label{fig.4}(Color online)
The conductivities $\sigma_{xx}$ (a) and $\sigma_{xy}$ (b) as functions of the gate voltage and magnetic
field for the gapless double quantum well, and traces of $\sigma_{xx}$ and
$\sigma_{xy}$ at $B=4$ T (c).}
\end{figure}

Now we turn to a detailed comparison between the experimental resistance plot of $R_{xx}(N_{s},B)$
and the theoretical LL spectrum. The fan chart in Fig. 5 (a) shows the LL energies $E^{(n)}_{i}$ as
functions of $B$ for the gapless DQW. Here $n$ is the LL index and $i$ is the subband index. Some of
the LLs are marked by the index $n=n_i$. The conduction band part consists of two sets of levels
originating from the E2 and H1 subbands, see Fig. 1 (b). The nonlinear dependence of LL energies
on $B$ is characteristic for relativistic Dirac particles. A striking feature of the LL spectrum
in DQWs is the presence of numerous crossings between the LLs. Some crossings of the lowest LLs are marked
by circles corresponding to the circles in the experimental plots. To increase resolution and visibility of the spectrum, we present
in Fig.5(b) the first derivative $dR_{xx}/dN_{s}$ of the measured longitudinal
resistance as a function of both the magnetic field $B$ and the density $N_{s}$, the negative sign of $N_{s}$
refers to the valence-band (hole) part of the energy spectrum. The filling factors corresponding to the minima
of $dR_{xx}/N_{s}$ (dark blue regions) are indicated. Figure 5 (c) shows the same in more details for magnetic
fields $0 < B < 4$ T. It is important to note that the LL spectra in both single and double QWs cannot be directly
accessed from the analysis of the $R_{xx}(B, N_{s})$ or $R_{xy}(B, N_{s})$ graphs because the slopes of the
stripes in Fig. 5 (b,c) are determined by the LL filling factors $\nu$, $d N_s/dB = \nu |e|/h c$, rather than
by the LL energy dependence. In contrast, the crossing points of the stripes in the plot $dR_{xx}/dN_{s}(B, N_{s})$
do correspond to the LL crossing. The knowledge of magnetic fields and carrier densities for these
crossings gives more information about energy spectra and makes it possible to determine parameters of the
DQWs with a higher accuracy. Thus, Fig. 5 (b,c) can be viewed as an experimental analogue of LL energy diagram.
It demonstrates a rich diamond-shaped pattern, which is absent in a single HgTe QW. The regular pattern
of LL crossing points is visible only in the conduction band part of the spectrum. The valence band
part has a dense and complicated spectrum containing many closely spaced LLs, which makes the HgTe QWs
essentially different from graphene, where electron-hole symmetry takes place in a wide interval of energies.
In the experiment, we are able to resolve only the $\nu=-1$ minimum in the valence band region.

\begin{figure}[ht!]
\includegraphics[width=9cm,clip=]{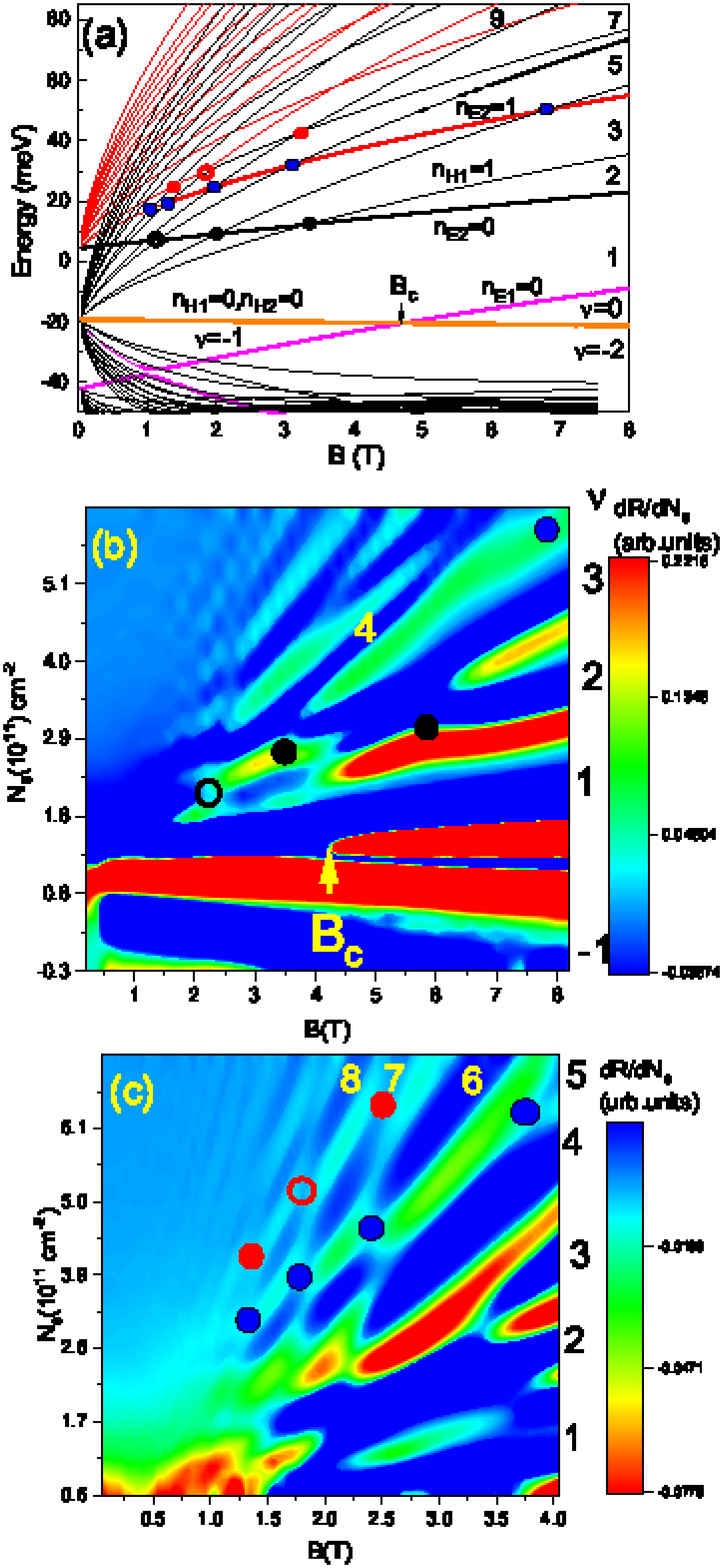}
\caption{\label{fig.5}(Color online) (a) Calculated energy spectrum
in the symmetric double quantum well with $d=6.5$ nm, $t=3$ nm, and $x_{Cd}=0.65$. The color map
of $dR(N_{s},B)/dN_{s}$ versus $N_{s}$ and $B$ at $T=4.2$ K is given for low (b) and high (c) filling
factors. Circles indicate LL crossing points, the empty circle stands for the anticrossing point.
The numbers at the right margin and inside the map represent the filling factors $\nu$.}
\end{figure}

\begin{figure}[ht!]
\includegraphics[width=9cm,clip=]{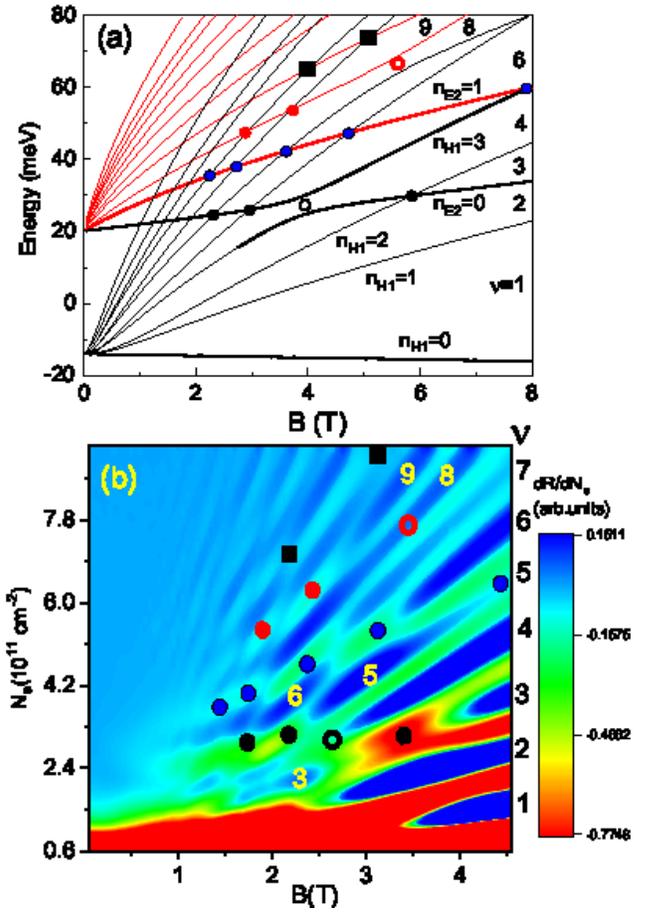}
\caption{\label{fig.6}(Color online) (a) Calculated energy spectrum (conduction band only)
in the biased (gapped) double quantum well with $d=6$ nm, $t=3$ nm, and $x_{Cd}=0.37$.
(b) The color map of $dR(N_{s},B)/dN_{s}$ versus $N_{s}$ and $B$ at $T=4.2$ K.
Circles indicate LL crossing points, the empty circle stands for the anticrossing point.
The numbers at the right margin and inside the map represent the filling factors $\nu$.}
\end{figure}

The spectrum in Fig. 5 (a) has four (in contrast to two in single QWs) special levels with zero LL index. Two
of them (marked by the indices $n_{H1}=0$ and $n_{H2}=0$) originate from the heavy hole subbands, do not
hybridize with other levels, and stay nearly degenerate at any $B$, following a linear dependence
$E^{(0)}_{H1,H2}= -\gamma B$. The splitting of $dR_{xx}/dN_{s}$ peak in Fig. 5 (b) near the
CNP at $B \equiv B_c \simeq 4$ T is identified with the crossing between these levels and zeroth LL of E1 subband
(the calculation gives $B_c =4.7$ T). The fact that the state with $\nu=-1$ at negative $N_s$ still persists at
$B > B_c$ can be explained if one assumes that application of the gate voltage not only shifts the density $N_s$ away
from zero but also creates an internal asymmetry of the DQW structure that opens a gap between H1 and H2 subbands.
The zero LLs of H1 and H2 subbands then have the spectra $E^{(0)}_{H1,H2}= -\gamma B \pm
\lambda |N_s|$, where $\lambda$ is roughly estimated as $2 \pi e^2(d+t)/\epsilon_0$ and equal to 4.3 meV per $10^{11}$ cm$^{-2}$
(here $\epsilon_0 \simeq 20$ is the static dielectric constant of HgTe and $d+t=9.5$ nm is the distance between the centers
of the wells). With increasing negative density, the Fermi level at high magnetic fields first becomes pinned
to $E^{(0)}_{H1}$ and then shifts down towards $E^{(0)}_{H2}$, staying either within the region $\nu=-1$ between these
two levels or pinned to $E^{(0)}_{H2}$ as long as $|N_s|/B < 0.5 \times 10^{11}$ cm$^{-2}$ T$^{-1}$ (note that the
capacitance of a single LL is approximately $0.25 \times 10^{11}$ cm$^{-2}$ per Tesla). This behavior of the Fermi
level is in accordance with the observed behavior of $R_{xx}$.

\begin{table}[ht]
\caption{\label{tab2} The magnetic fields and the filling factors at which the LL crossings
occur for the gapless DQW. The index $\star$ denotes LL anticrossing.
The fields $B^*_{theor}$ are calculated for $t=2.6$ nm.}
\begin{ruledtabular}
\begin{tabular}{lccccc}
$n_{E2}$ & $n_{H1}$ & $\nu$ & $B_{exp}$ (T) &  $B_{theor}$ (T) &  $B^*_{theor}$ (T)\\
\hline
0&   1 & 2 & 5.6 & 3.4 &                  5.7 \\
\hline
0&   2 & 3 & 3.2 & 2.0 &                  3.0 \\
\hline
0&   3 & 4 & 2.2 & 1.2$^{\star}$ & 1.8$^{\star}$ \\
\hline
1&   2 & 4 & 7.5 & 6.8 &                  8.6 \\
\hline
1&   3 & 5 & 3.8 & 3.1 &                  4.3 \\
\hline
1&   4 &6 & 2.3 & 2.0 &                  2.7 \\
\hline
1&   5 & 7 & 1.8 & 1.3 &                  1.9 \\
\hline
1&   6 & 8 & 1.35 & 1.05 &       1.8 \\
\hline
2&   4 & 7 & 2.5 & 3.25 &   4.5 \\
\hline
2&   5 & 8 & 1.8 & 1.85$^{\star}$ &   2.7$^{\star}$ \\
\hline
2&   6 & 9 & 1.37 & 1.35 &                  1.9 \\
\hline
\end{tabular}
\end{ruledtabular}
\end{table}

\begin{table}[ht]
\caption{\label{tab3} The magnetic field values and the filling factors at which the LL crossings
occur for the gapped DQW. The index $\star$ denotes LL anticrossing.}
\begin{ruledtabular}
\begin{tabular}{lccccccc}
$n_{E2}$ & $n_{H1}$ & $\nu$ & $B_{exp}$ (T) &  $B_{theor}$ (T) \\
\hline
0&   2 & 3 & 3.5 & 5.8 \\
\hline
0&   3 & 4 & 2.6 & $3.9^{\star}$\\
\hline
0&   4 & 5 & 2.2 & 2.9 \\
\hline
0&   5 & 6 & 1.7 & 2.3 \\
\hline
1&   3 & 5& 4.5 & 7.9 \\
\hline
1&   4 & 6& 3.2 & 4.7 \\
\hline
1&   5 & 7& 2.5 & 3.6 \\
\hline
1&   6 & 8& 1.8 & 2.7 \\
\hline
1&   7 & 9& 1.45 & 2.25 \\
\hline
2&   5 & 8 & $3.3^{\star}$ & $5.5^{\star}$ &  \\
\hline
2&   6 & 9 & 2.4 & 3.7 &  \\
\hline
2&   7 & 10 & 1.9 & 2.9 &  \\
\hline
3&   6 & 10 & 3.1 & 5.05 &  \\
\hline
3&   7 & 11 & 2.15 & 4.05 &  \\

\hline
\end{tabular}
\end{ruledtabular}
\end{table}

Figure 6 shows the theoretical LL spectra and experimental $dR_{xx}/dN_{s}(B, N_{s})$ map for the DQW structure with $d=6$ nm.
Since the resistance behavior in zero magnetic field reveals the presence of a gap $\Delta=11$ meV
at CNP, we have calculated the spectra by assuming an asymmetry caused by the electrostatic potential $eEz$
with the field $E=13$ kV/cm, which produces $\Delta=11$ meV at $B=0$. This leads to splitting of the zero LLs by
$E^{(0)}_{H1}-E^{(0)}_{H2} \simeq \Delta$ already at $N_s=0$, so the conduction and the valence bands
are separated. Here we describe the LLs only in the conduction band. Except for the shift of the subbands
and crossing points, the picture of the spectrum is similar to that of Fig. 5 (a). However, the
electric field leads to a superlinear dispersion of the H1 subband LL energies $E^{(n)}_{H1}(B)$ at small
$B$, while in asymmetric DQWs with larger $t$ and $x_{Cd}$ the dispersion can even become non-monotonic. The
resistance map again demonstrates a rich diamond-shaped pattern that qualitatively correlates with the
calculated fan chart. Notice that the experimental map does not show a well-defined resistance maximum in
the place where the anticrossing of LLs due to tunnel coupling is expected from the theory (open circle), in
contrast to the places of true crossing (filled circles), and a similar behavior is seen in Fig. 5 (b),
confirming the consistence between experiment and theory.

The positions of some important LL crossing points for both DQWs are summarized in the Tables II and III.
Though the theory allows us to identify each of the main crossings observed, there is a lack
of quantitative agreement between experimental and theoretical values of the corresponding magnetic fields.
One of the reasons of this disagreement is the presence of the transverse field induced by the gate, which
is not taken into account in the calculations. Since this field depends on the filling factor, a more
precise description of LL spectrum requires a self-consistent approach, which is beyond the scope of
this paper. Another important reason is related to possible measurement errors in determination of well widths
and barrier thicknesses by the ellipsometry method. Indeed, the positions of crossing points, especially the
ones involving $n_{E2}=0$ level, are strongly sensitive to the barrier thickness $t$, because a variation of
$t$ strongly changes the tunnel-hybridization gap between E1 and E2 states thereby affecting the relative position
of E2 subband with respect to H1 one. In particular, taking $t=2.6$ nm, which corresponds to four monolayers of
Cd$_{0.65}$Hg$_{0.35}$Te, we obtain a better agreement between experiment and theory for the gapless DQW, as shown
in Table 2. Finally, one should mention limitations of Kane model in application to heterostructures with narrow
(several monolayers) potential barriers. A lack of quantitative agreement between experiment and theory
in determination of LL positions in HgTe DQWs has been also pointed out by the other authors \cite{bovkun}.

\section{Conclusions}

We have studied transport properties of the 2D fermion systems in HgTe-based DQWs in both gapless semimetal
and gapped topological insulator phases in the presence of a perpendicular magnetic field. By changing
independently the electron density and the magnetic field, we have observed multiple crossing of LLs
originating from different subbands, according to earlier theoretical predictions \cite{kristopenko}.
The data indicates a similarity of the energy spectrum of gapless HgTe DQWs to that of bilayer graphene.
However, in contrast to bilayer graphene, the separation of subbands in such structures is small, which
allowed us to observe numerous crossigns at relatively low densities. In the charge neutrality point, we
have indicated a special crossing of zeroth LLs, which is a hallmark of topologically non-trivial
materials described by the Dirac-like BHZ Hamiltonian \cite{konig}. The pattern of the crossing points
in both kinds of HgTe DQWs generally correlates with the theoretical picture of energy spectrum of LLs.
The reasons for deviations of the experimental crossing points from the theoretical ones are discussed.

It is expected that a symmetric HgTe DQW in the gapless semimetal phase can be switched electrically to the
2D topological insulator state as a gap between heavy hole subbands H1 and H2 opens in the presence of a
transverse electric field. By observing the behavior of the resistance at low filling factors, we have obtained
an indirect evidence that application of the bias to a single (top) gate not only changes the density but also
opens the gap. An independent control of the energy spectrum and electron density can possibly be achieved with
the use of double gating technology, which is still missing for MBE-grown HgTe DQWs. This may become useful for
a more precise evaluation of the parameters which determine the energy spectrum. The double gating can also enable
the exploration of the quantum Hall ferromagnetism in Dirac materials and development of electrically tunable
topological insulators in multilayer HgTe-based systems.

\section*{ACKNOWLEDGEMENTS}

The financial support of this work by RFBI Grant No. 18-02-00248a., FAPESP (Brazil), and CNPq (Brazil) is acknowledged.


\end{document}